\documentclass[twocolumn,amsmath,amssymb,10pt,superscriptaddress,a4paper,letterpaper,fleqn]{revtex4-1}
              \usepackage{amssymb}
              \usepackage{epsfig}
              \usepackage{graphicx}
              \usepackage{dcolumn}
              \usepackage{array}
              \usepackage{bm}
              \usepackage{fancyheadings}
              \usepackage{longtable}
              \usepackage{multirow}
              \usepackage{float}
              \usepackage{afterpage}  
              \usepackage{color}
              
              \setlongtables
              \usepackage[breaklinks=true,linkbordercolor={1 1 1}]{hyperref}
              
\begin{document}
              \setcounter{page}{1}
              
              \title{
              \qquad \\ \qquad \\ \qquad \\  \qquad \\  \qquad \\ \qquad \\ 
              Correlations between charge radii, E0 transitions, and M1 strength}
              
              \author{P.~Van~Isacker}
              \email[Corresponding author:\\ ]{isacker@ganil.fr}
              \affiliation{Grand Acc\'el\'erateur National d'Ions Lourds, CEA/DSM-CNRS/IN2P3,
              BP 55027, F-14076 Caen Cedex 5, France}

              \date{\today} 
              
              \begin{abstract}
              In the framework of the interacting boson model,
              relations are derived 
              between nuclear charge radii, electric monopole transition rates,
              and summed magnetic dipole transition in even-even nuclei.
              The proposed correlations are tested in the rare-earth region.
              \end{abstract}
              \maketitle
              
              
              \lhead{ND 2013 Article $\dots$}
              \chead{NUCLEAR DATA SHEETS}
              \rhead{P.~Van~Isacker}
              \lfoot{}
              \rfoot{}
              \renewcommand{\footrulewidth}{0.4pt}

              \section{Introduction}
              \label{s_intro}
              In recent papers~\cite{Zerguine08,Zerguine12}
              a simultaneous description was proposed
              of charge radii and electric monopole transitions
              of nuclei in the rare-earth region.
              The purpose of the studies was to examine to what extent
              a purely collective interpretation of nuclear $0^+$ levels
              is capable of yielding a coherent and consistent picture of both properties.
              
              In this contribution a further correlation is proposed
              between both of the above properties
              and summed M1 strength as observed in even-even nuclei.
              The framework used to establish this correlation
              is the interacting boson model (IBM) of Arima and Iachello~\cite{Iachello87}
              which describes nuclear collective excitations
              in terms of $s$ and $d$ bosons
              with angular momenta $\ell=0$ and 2, respectively.
              The simplest version of the model is used, \mbox{IBM-1},
              which makes no distinction between neutron and proton bosons.
              
              A brief recall of the necessary operators is given in Sect.~\ref{s_operators}
              and previous results~\cite{Zerguine12} on isotope shifts
              are summarized in Sect.~\ref{s_dr2}.
              Correlations between the summed M1 strength
              and isotope shifts, isomer shifts, and $\rho({\rm E0})$ values
              are pointed out in Sects.~\ref{s_m1ist}, \ref{s_m1ism}, and~\ref{s_m1rho}, respectively.
              Finally, topics for further study are listed in Sect.~\ref{s_look}.
              
              \section{Operators in the \mbox{IBM-1}}
              \label{s_operators}
              In the \mbox{IBM-1} the charge radius operator
              is taken as the most general scalar expression,
              linear in the generators of U(6)~\cite{Iachello87},
              \begin{equation}
              \hat T(r^2)=
              \langle r^2\rangle_{\rm c}+
              \alpha N_{\rm b}+
              \eta\frac{\hat n_d}{N_{\rm b}},
              \label{e_r2}
              \end{equation}
              where $N_{\rm b}$ is the total number of $s$ and $d$ bosons,
              $\hat n_d$ is the $d$-boson number operator,
              and $\alpha$ and $\eta$ are coefficients with units of length$^2$.
              The first term in Eq.~(\ref{e_r2}), $\langle r^2\rangle_{\rm c}$,
              is the square of the charge radius of the core nucleus.
              The second term accounts for the (locally linear) increase
              in the charge radius due to the addition of two nucleons.
              The third term in Eq.~(\ref{e_r2})
              stands for the contribution to the charge radius due to deformation.
              It is identical to the one given in Ref.~\cite{Iachello87}
              but for the factor $1/N_{\rm b}$.
              This factor is included here
              because it is the {\em fraction} $\langle\hat n_d\rangle/N_{\rm b}$
              which is a measure of the quadrupole deformation
              ($\beta^2$ in the geometric collective model)
              rather than the matrix element $\langle\hat n_d\rangle$ itself.
              
              Two quantities can be derived from charge radii:
              isotope and isomer shifts.
              The former measures the difference in charge radius of neighboring isotopes.
              For the difference between even-even isotopes
              one finds from Eq.~(\ref{e_r2}) 
              \begin{eqnarray}
              \Delta \langle r^2\rangle&\equiv&
              \langle r^2\rangle_{0_1^+}^{A+2}-\langle r^2\rangle_{0_1^+}^A
              \nonumber\\&=&
              |\alpha|+\frac{\eta}{N_{\rm b}}
              \left(\langle\hat n_d\rangle^{A+2}_{0_1^+}-
              \langle\hat n_d\rangle^A_{0_1^+}\right),
              \label{e_ips}
              \end{eqnarray}
              where $\langle\hat n_d\rangle_{J^\pi}$
              is a short-hand notation for $\langle J^\pi|\hat n_d|J^\pi\rangle$.
              Isomer shifts are a measure of the difference in charge radius
              between an excited ({\it e.g.}, the $2_1^+$) state
              and the ground state, and are given by
              \begin{eqnarray}
              \delta\langle r^2\rangle&\equiv&
              \langle r^2\rangle^A_{2_1^+}-\langle r^2\rangle^A_{0_1^+}
              \nonumber\\&=&
              \frac{\eta}{N_{\rm b}}
              \left(\langle\hat n_d\rangle^{A+2}_{2_1^+}-
              \langle\hat n_d\rangle^A_{2_1^+}\right).
              \label{e_ims}
              \end{eqnarray}
              
              Once the form of the charge radius operator is determined,
              the E0 transition operator follows from the relation~\cite{Zerguine08,Zerguine12}
              \begin{equation}
              \hat T({\rm E0})=
              (e_{\rm n}N+e_{\rm p}Z)\hat T(r^2),
              \label{e_e0}
              \end{equation}
              where $e_{\rm n}$ ($e_{\rm p}$) is the neutron (proton) effective charge.
              Since for E0 transitions the initial and final states are different,
              neither the constant $\langle r^2\rangle_{\rm c}$
              nor $\alpha N_{\rm b}$ in Eq.~(\ref{e_r2}) contribute to the transition,
              and the $\rho({\rm E0})$ value equals
              \begin{equation}
              \rho({\rm E0};i\rightarrow f)=
              \frac{e_{\rm n}N+e_{\rm p}Z}{eR^2}
              \frac{\eta}{N_{\rm b}}
              |\langle f|\hat n_d|i\rangle|.
              \label{e_rho}
              \end{equation}
              
              The magnetic dipole operator in the \mbox{IBM-1}
              is of the form~\cite{Iachello87}
              \begin{equation}
              \hat T({\rm M1})=
              \sqrt{\frac{3}{4\pi}}
              \left(g_\nu\hat L_\nu+g_\pi\hat L_\pi\right),
              \label{e_m1}
              \end{equation}
              where $\hat L_\nu$ ($\hat L_\pi$)
              is the angular momentum operator for the neutrons (protons)
              and $g_\nu$ ($g_\pi$) the $g$ factor of the neutron (proton) boson.
              
              \section{Isotope shifts}
              \label{s_dr2}
              To test the relation between charge radii and E0 transitions,
              a systematic study of even-even isotopes
              from Ce ($Z=58$) to W ($Z=74$) was carried out in Ref.~\cite{Zerguine12}.
              The analysis required the knowledge of structural information
              concerning the ground and excited levels
              to which an \mbox{IBM-1} Hamiltonian is adjusted.
              A general one- and two-body Hamiltonian is taken
              with six parameters which are constant for a given isotope series,
              except one which is allowed to vary with valence neutron and proton numbers. 
              Details can be found in Ref.~\cite{Zerguine12}.
              
              \begin{figure}
              \includegraphics[width=1\columnwidth]{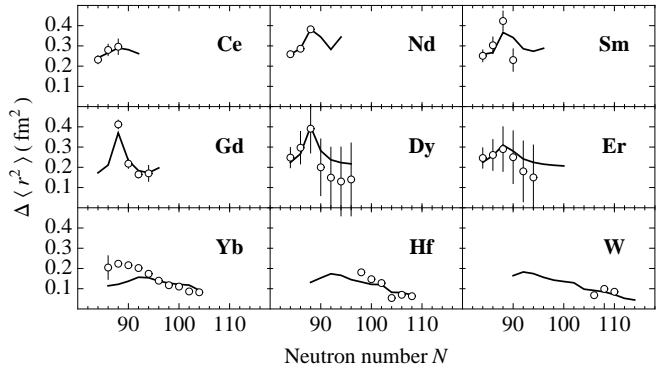}
              \caption{
              Experimental (points) and calculated (lines) isotope shifts $\Delta\langle r^2\rangle$,
              for isotopic chains in the rare-earth region from Ce to W.
              The data are taken
              from Ref.~\cite{Cheal03} for Ce,
              from Ref.~\cite{Otten89} for Nd, Sm, Dy, Er, and Yb,
              from Ref.~\cite{Fricke95} for Gd,
              from Ref.~\cite{Angeli04} for Hf,
              and from Ref.~\cite{Jin94} for W.}
              \label{f_dr2}
              \end{figure}
              Once the \mbox{IBM-1} Hamiltonian is obtained for a given nucleus,
              matrix elements of the operators discussed in Sect.~\ref{s_operators}
              depend solely on the coefficients appearing in the operators.
              Isotope shifts $\Delta\langle r^2\rangle$, according to Eq.~(\ref{e_ips}),
              depend on the coefficients $|\alpha|$ and $\eta$.
              The coefficient $|\alpha|$ is adjusted
              for each isotope series separately,
              while $\eta$ is kept constant for all isotopes,
              $\eta=0.50$~fm$^2$.
              The resulting isotope shifts are shown in Fig.~\ref{f_dr2}.
              The largest peaks in the isotope shifts
              occur for $^{152-150}$Sm, $^{154-152}$Gd, and $^{156-154}$Dy,
              that is, for the difference in radii between $N=90$ and $N=88$ isotopes.
              The peak is smaller below $Z=62$ for Ce and Nd,
              and fades away above $Z=66$ for Er, Yb, Hf, and W.
               
              \section{Correlation between M1 strength and isotope shifts}
              \label{s_m1ist}
              It is known from the work of Ginocchio~\cite{Ginocchio91}
              that the summed M1 strength from the ground state to the scissors mode
              (for a review on the latter, see Ref.~\cite{Heyde10})
              is related to the ground-state matrix element
              of the $d$-boson number operator $\hat n_d$,
              \begin{eqnarray}
              S({\rm M1};0^+_1)&\equiv&
              \sum_f
              B({\rm M1};0^+_1\rightarrow1^+_f)
              \nonumber\\&=&
              \frac{3}{4\pi}(g_\nu-g_\pi)^2
              \frac{6N_\nu N_\pi}{N_{\rm b}(N_{\rm b}-1)}
              \langle\hat n_d\rangle_{0^+_1},
              \label{e_bm1}
              \end{eqnarray}
              where $N_\nu$ ($N_\pi$) is the number of neutron (proton) bosons
              and the sum is over all possible $1^+$ states
              characterized by the label $f$.
              
              To establish a connection between summed M1 strength and isotope shifts,
              one rewrites the relation~(\ref{e_bm1}) as follows:
              \begin{eqnarray}
              \tilde S({\rm M1};0^+_1)&\equiv&
              \frac{N_{\rm b}-1}{N_\nu}S({\rm M1};0^+_1)
              \nonumber\\&=&
              \frac{9}{2\pi}(g_\nu-g_\pi)^2\frac{N_\pi}{N_{\rm b}}
              \langle\hat n_d\rangle_{0^+_1}.
              \label{e_bm1r}
              \end{eqnarray}
              The rewritten relation is such
              that all $N$-dependent quantities
              ({\it i.e.}, $N_{\rm b}$ and $N_\nu$)
              are shifted to the left-hand side of the equation,
              except for the factor $1/N_{\rm b}$
              which precisely coincides with the $N$ dependence
              as it appears in the definition of the isotope shifts~(\ref{e_ips}).
              The tilde in $\tilde S({\rm M1};0^+_1)$ is used as a reminder
              that it is not the summed M1 strength
              but rather the summed M1 strength weighted by an $N$-dependent factor.
              One now defines the difference
              \begin{eqnarray}
              \lefteqn{\Delta\tilde S({\rm M1})\equiv
              \tilde S({\rm M1};0^+_1)^{A+2}-
              \tilde S({\rm M1};0^+_1)^A}
              \nonumber\\&&=
              \frac{9}{2\pi}(g_\nu-g_\pi)^2\frac{N_\pi}{N_{\rm b}}
              \left(\langle\hat n_d\rangle^{A+2}_{0_1^+}-
              \langle\hat n_d\rangle^A_{0_1^+}\right).
              \label{e_bm1d}
              \end{eqnarray}
              The quantity between brackets on the right-hand side of Eq.~(\ref{e_bm1d})
              is precisely the one that occurs in the expression~(\ref{e_ips}) for the isotope shift
              and hence the following relation is established:
              \begin{equation}
              \Delta\tilde S({\rm M1})=
              \frac{9}{2\pi}\frac{(g_\nu-g_\pi)^2}{\eta}N_\pi
              \left(\Delta \langle r^2\rangle-|\alpha|\right).
              \label{e_bm1ips}
              \end{equation}
              
              This relation is different
              from the one proposed by Heyde {\it et al.}~\cite{Heyde93}
              which correlates summed M1 strength and charge radii themselves.
              This has two drawbacks:
              (i) the quantity $\langle r^2\rangle_{\rm c}$,
              which appears in the operator~(\ref{e_r2}), is ill determined
              and (ii) charge radii are not that well known
              as their differences (isotope shifts).
              The relation~(\ref{e_bm1ips}) is model dependent
              since it involves the coefficient $\alpha$
              which cannot be neglected (it represents an important part of the isotope shift)
              and which varies from one isotope series to another.
              Nevertheless, with the values for $\alpha$
              obtained from the isotope shifts (see Sect.~\ref{s_dr2}),
              the relation~(\ref{e_bm1ips}) can be tested in the rare-earth region.
              The slope in the correlation plot
              depends on the constants $\eta$ and $g_\nu-g_\pi$.
              The former is known from the isotope shifts
              while the latter can be obtained by adjusting the expression~(\ref{e_bm1})
              to the observed summed M1 strength in rare-earth nuclei~\cite{Pietralla95,Enders05},
              leading to $|g_\nu-g_\pi|\approx0.83$~$\mu_{\rm N}$.
              
              \begin{figure}
              \includegraphics[width=0.95\columnwidth]{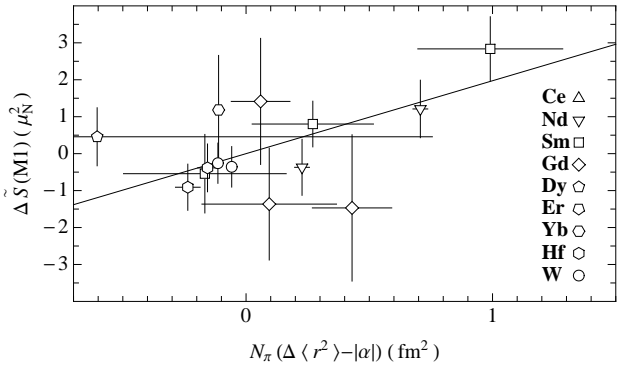}
              \caption{Test of the relation~(\ref{e_bm1ips})
              between differences of summed M1 strength in neighboring isotopes
              and isotope shifts.
              The summed M1 strength is taken from Ref.~\cite{Enders05}.
              The slope of the line is determined by $\eta$ and $g_\nu-g_\pi$,
              coefficients that appear in the charge radius and M1 operators, respectively.}
              \label{f_bm1r2}
              \end{figure}
              The resulting correlation plot is shown in Fig.~\ref{f_bm1r2}.
              Two compilations exist of the observed summed M1 strength,
              one by Pietralla {\it et al.}~\cite{Pietralla95}
              and a second by Enders {\it et al.}~\cite{Enders05}.
              They give broadly consistent results
              and hence lead to similar correlation plots.
              The latter, more recent compilation is taken here.
              Large error bars on both quantities
              preclude at the moment a conclusive test of the proposed correlation.
              In addition, there is an uncertainty (not included in Fig.~\ref{f_bm1r2})
              associated with the coefficient $\alpha$.
              The large errors follow from poorly determined isotope shifts
              ({\it e.g.}, in dysprosium)
              but also because {\em differences} of summed M1 strength should be considered
              and not the summed M1 strength itself.
              
              An alternative strategy, to be explored in the future,
              is to use the relation~(\ref{e_bm1ips})
              to fix the coefficients $\alpha$ from the experimental summed M1 strength
              and use the resulting values in the calculation of nuclear radii.
              
              \section{Correlation between M1 strength and isomer shifts}
              \label{s_m1ism}
              The relation~(\ref{e_bm1}) has been generalized to summed M1 strength
              from an arbitrary state to the scissors mode
              built on top of that state~\cite{Ginocchio97,Smirnova02},
              \begin{widetext}
              \begin{equation}
              S({\rm M1};J_i)\equiv
              \sum_{f(\neq i)}
              B({\rm M1};J_i\rightarrow J_f)=
              \frac{3}{4\pi}(g_\nu-g_\pi)^2
              \frac{6N_\nu N_\pi}{N_{\rm b}(N_{\rm b}-1)}
              \left(\langle\hat n_d\rangle_{J_i}-
              \frac{J_i(J_i+1)}{6N_{\rm b}}\right),
              \label{e_bm1g}
              \end{equation}
              where the initial state $i$ is excluded from the sum over $f$.
              The difference 
              \begin{equation}
              \delta S({\rm M1})\equiv
              S({\rm M1};2^+_1)-S({\rm M1};0^+_1)=
              \frac{3}{4\pi}(g_\nu-g_\pi)^2
              \frac{6N_\nu N_\pi}{N_{\rm b}(N_{\rm b}-1)}
              \left(\langle\hat n_d\rangle_{2_1^+}-\frac{1}{N_{\rm b}}-
              \langle\hat n_d\rangle_{0_1^+}\right),
              \label{e_bm1gd}
              \end{equation}
              \end{widetext}
              contains the same combination of matrix elements of $\hat n_d$
              as the one that appears in the isomer shift~(\ref{e_ims}),
              and therefore the following relation is established:
              \begin{equation}
              \delta S({\rm M1})=
              \frac{9}{2\pi}\frac{(g_\nu-g_\pi)^2}{\eta}
              \frac{N_\nu N_\pi}{N_{\rm b}-1}
              \left(\delta \langle r^2\rangle-\frac{\eta}{N_{\rm b}^2}\right).
              \label{e_bm1ims}
              \end{equation}
              Some isomer shifts are known in the rare-earth region;
              the data are more than 30 years old and often discrepant.
              Nothing is known about M1 strength built on excited states.
              The relation~(\ref{e_bm1ims}) therefore remains untested.
              
              \section{Correlation between M1 strength and $\rho$(E0) values}
              \label{s_m1rho}
              The E0 operator is directly proportional to $\hat n_d$,
              unlike the charge radius operator
              which involves additional terms
              which complicate the relation
              between the summed M1 strength and charge radii,
              as shown above.
              On the other hand, the matrix element of $\hat n_d$
              appearing in the sum rule~(\ref{e_bm1}) is diagonal
              while a $\rho({\rm E0})$ value
              involves a non-diagonal matrix element of $\hat n_d$.
              A relation between the two matrix elements
              can nevertheless be obtained in the symmetry limits of the \mbox{IBM-1}.
              In particular, in the SU(3) limit, appropriate for deformed nuclei,
              the following analytic expressions are found~\cite{Subber88}:
              \begin{eqnarray}
              \langle0^+_1|\hat n_d|0^+_1\rangle&=&
              \frac{4N_{\rm b}(N_{\rm b}-1)}{3(2N_{\rm b}-1)},
              \label{e_su3mes}\\
              |\langle0^+_\beta|\hat n_d|0^+_1\rangle|&=&
              \left[\frac{8(N_{\rm b}-1)^2N_{\rm b}(2N_{\rm b}+1)}
              {9(2N_{\rm b}-3)(2N_{\rm b}-1)^2}\right]^{1/2},
              \nonumber
              \end{eqnarray}
              where $0^+_\beta$ is the second $0^+$ level
              as calculated in the \mbox{IBM-1}.
              It should be pointed out that
              the beta-vibrational state $0^+_\beta$, if it exists at all in nuclei,
              is not necessarily the observed $0^+_2$ level
              since non-collective excitations might occur at a lower energy.
              From the expressions~(\ref{e_su3mes}) the ratio of matrix elements can be derived,
              resulting in the following relation,
              valid in the large-$N_{\rm b}$ limit:
              \begin{eqnarray}
              B({\rm M}1;0^+_1\rightarrow1^+_1)&\approx&
              \frac{9}{\pi}(g_\nu-g_\pi)^2
              g(N,Z,N_\nu,N_\pi)
              \nonumber\\&&\times
              \frac{r_0^2}{\eta}
              \rho({\rm E}0;0^+_\beta\rightarrow0^+_1),
              \label{e_bm1rho}
              \end{eqnarray}
              where $g(N,Z,N_\nu,N_\pi)$ is the function
              \begin{equation}
              g(N,Z,N_\nu,N_\pi)=
              \frac{e(N+Z)^{2/3}}{e_{\rm n}N+e_{\rm p}Z}
              \frac{N_\nu N_\pi}{\sqrt{2N_{\rm b}}},
              \label{e_gfun}
              \end{equation}
              and $r_0$ is the constant that appears in the radius parameterization $R=r_0A^{1/3}$.
              In the SU(3) limit only one $1^+$ state is excited
              and only the $\beta$-vibrational state decays by E0 to the ground state,
              as is indicated in Eq.~(\ref{e_bm1rho}).
              
              \begin{figure}
              \includegraphics[width=0.95\columnwidth]{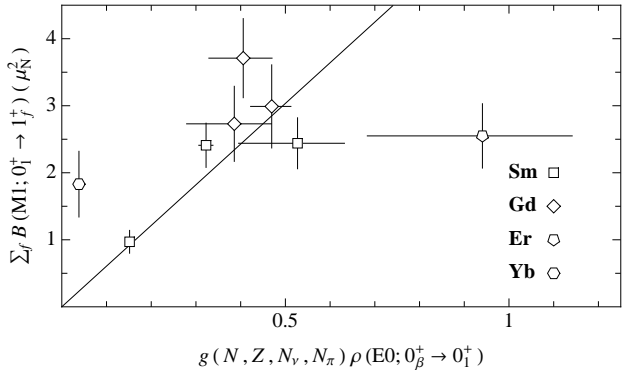}
              \caption{Test of the relation~(\ref{e_bm1rho})
              between summed M1 strength
              and $\rho({\rm E}0;0^+_\beta\rightarrow0^+_1)$ values
              multiplied with the function $g$ defined in Eq.~(\ref{e_gfun}).
              The summed M1 strength is taken from Ref.~\cite{Enders05}.
              The slope of the line is determined by $\eta$, $r_0^2$, and $g_\nu-g_\pi$,
              coefficients that are obtained from a fit to radii, E0 and M1 strength.}
              \label{f_bm1rho}
              \end{figure}
              The relation~(\ref{e_bm1rho}) is valid only in the SU(3) limit
              which might jeopardize its use in transitional nuclei.
              One may nevertheless attempt to apply it to the entire rare-earth region.
              The ratio $r_0^2/\eta=3.08$
              and the effective charges $e_{\rm n}=0.5e$ and $e_{\rm p}=e$
              are determined from a fit to radii~\cite{Zerguine12}.
              The correlation~(\ref{e_bm1rho}) can now be tested (see Fig.~\ref{f_bm1rho})
              for the eight nuclei in the rare-earth region
              where both E0 and M1 properties are known
              ($^{150,152,154}$Sm, $^{154,156,158}$Gd, $^{166}$Er, and $^{172}$Yb).
              The $^{172}$Yb point is conspicuously off the line
              which calls for a search for E0 strength in this nucleus.
              The $^{166}$Er point follows from a recent experiment~\cite{Wimmer09} 
              where the {\em fourth} $J^\pi=0^+$ level at 1934~keV
              has been identified as the band head of the $\beta$-vibrational band
              with a sizable E0 matrix element to the ground state.
              
              \section{Outlook}
              \label{s_look}
              This work identified the following three topics for further study.
              The analysis of the correlation between summed M1 strength
              and $\rho({\rm E0})$ values
              should be extended to transitions between $J^\pi\neq0^+$ states.
              Secondly, the relation~(\ref{e_bm1rho})
              should be generalized to transitional nuclei.
              Finally, on the experimental front,
              there is crying need for precise data
              on isotope shifts through the shape transition in rare-earth nuclei.\\
              
              \section*{Acknowledgments}
              Thanks are due to Salima Zerguine and Abdelhamid Bouldjedri
              in collaboration with whom part of this work was done.
              Conversations and exchanges with
              Joe Ginocchio, Norbert Pietralla, Achim Richter, and Peter von Neumann-Cosel
              are gratefully acknowledged.

              \end{document}